# 360-degree fringe-projection profilometry of discontinuous solids with 2 co-phased projectors and 1-camera

**Manuel Servin,**[*] **Guillermo Garnica and J. M. Padilla.**

*Centro de Investigaciones en Optica A. C., 115 Loma del Bosque,
Col. Lomas del Campestre, 37150 Leon Guanajuato, Mexico.*
*November 2014*
[*]*mservin@cio.mx*

**Abstract:** We describe a theoretical analysis and experimental set-up of a co-phased 360-degree fringe-projection profilometer. This 360-degree profilometer is built using 2-projections and 1-camera and can digitize discontinuous solids with diffuse light surface. A 360-degree profilometer rotate the object a full revolution to digitize the analyzing solid. Although 360-degree profilometry is not new, we are proposing however a new experimental set-up which permits the 360-degree digitalization of discontinuous (piecewise-continuous) solids. The main advantage of using this co-phased 2-projectors profilometer is that self-occluding shadows due to discontinuities are solved efficiently. Previous 1-projector, 1-camera 360-degree profilometers generate self-occluding shadows at the solid discontinuities. Yet another advantage of our new profilometer is a trivial line-by-line fringe-data assembling from all 360-degree perspectives. Finally we used a 400 steps/rotation turntable, and a 640x480 pixels CCD camera. Higher resolutions and less-noisy phase demodulation are trivial by increasing the angular-resolution and phase-step number without any change on our co-phased profilometer. This profilometer may be used to digitize complex real life solids for possible 3D printing. A previous preliminary report of this work was published in the arXiv.org repository http://arxiv.org/ftp/arxiv/papers/1408/1408.6463.pdf .

## References and links

**1. Introduction**

Fringe projection profilometry using phase-demodulation techniques is well known since the classical paper by Takeda et al. in 1982 [1]. Although this phase-measuring profilometry technique effectively demonstrated that 3D digitalization was possible using a single carrier fringe pattern, it cannot digitize the 360-degree full view of the solid.  One needs to position the solid over a turntable to have access to solid perspective from all 360-degree perspectives. As far as we know the first researcher to implement an automated 360-degree profilometer was Halioua et al. in 1985 [2]. Halioua used a grating projector along with a 3-step phase shifter set-up and a turntable to obtain the 360-degree profilometry of a human mannequin head [2].  Later on in 1991 Cheng et al. have also positioned the 3D object in a turntable in order to rotate it 360-degree and obtain the full object and projected over it a carrier frequency fringe-pattern [3]. Also widely used, are striped-light projection profilometers. These single line-light profilometers normally use intensity triangulation-based methods which are less accurate than phase-demodulation methods. Asundi published an interesting technique based on a striped-light projection for 360-degree profilometry [4]. Also Chang et al. used a light-stripe from a laser diode [5] and automatically reconstructed a solid with 360 degrees using a neural network. Gomes et al. used a projected linear-grating to study the human trunk looking for spinal deformities; they used Fourier profilometry for their purpose [6]. Later on Song et al. used a fringe grating projector and phase-shifting interferometry of a rotating object for 360º degree profilometry [7]. Asundi et al. also implemented a fast 360-degree profilometer using a time delay integration imaging for solid digitalization [8]. The state of the art on 3D profilometry was reviewed in 2001 by Sue and Chen but they have just included a single paper of 360-degree profilometry [9]. More recently Zhang et al. have used 360-degree profilometry for flow analysis in mechanical engines [10]. In 2005 Munoz-Rodriguez et al. used triangulation for 3D object reconstruction by projecting a stripe-light and Hu moments [11]. In 2008 Trujillo-Shiaffino et al. used 3D profilometry based on a single line projection and triangulation of a smooth rotating-symmetric object [12]. More recently Shi et al. used 360-degree fringe pattern projection profilometry applied to fluorescent molecular tomography [13]. Some researchers have used shearing interferometry to project high quality linear fringes for 360 degree profilometry [14]. In this paper we are not discussing calibration issues, because we have not used any new or non-standard calibration strategy apart from those well-known in 3D profilometry [15]. Also we have not used any new or non-standard phase unwrapping algorithm [14,15]. Given the low noise of the white-light projected fringes and the noise-rejection of the 4-step least-squares phase-shifting demodulation, we have unwrapped our phase maps using standard flood-fill line-integration unwrapping. From this small review we see that 360-degree profilometry is already a mature research field and applications in industrial inspection and robotic vision are well known. But previous efforts [2-15] have mainly concentrated in digitizing very smooth, quasi-cylindrical objects. This is because smooth, quasi-cylindrical solids are easier to digitize with previous 360-degree profilometry [2-15].

The co-phased 2-projectors, single-camera fringe-projection profilometer herein presented is a blending of two previous published techniques by the same authors [17, 18]. We have tested our proposed 2-projector co-phased technique digitizing a far more complicated solid than previously reported [2-18]; we have digitized a human skull model all around it (almost $4\pi$-steradians). The plastic skull model is a discontinuous non-convex 3D-solid and it is certainly not a quasi-cylindrical object. By digitizing this skull we demonstrate that our co-phased 2-projections, 1-camera 360-degree technique is more robust and can digitize more complex solids than previous 360-degree fringe-projection profilometers [2-18]. This paper is an extended and more detailed and extended presentation of our previous paper http://arxiv.org/ftp/arxiv/papers/1408/1408.6463.pdf .

## 2. Co-phased fringe-projection profilometry using 2-projections and 1-camera

Here we review the experimental set-up basis for a co-phased profilometer having 2-projectors and 1-camera. Let us begin with co-phase profilometry (see Fig. 1) [17]. Figure 1 shows the co-phased profilometer when linear fringes are projected over a half sphere [17]. The left ($L$) and right ($R$) projected fringes as seen at the CCD camera plane are given by,

$$I_L(x,y,\alpha) = a_L(x,y) + b_L(x,y)\cos[\omega_0 x - g\,z(x,y) + \alpha],$$
$$I_R(x,y,\alpha) = a_R(x,y) + b_R(x,y)\cos[\omega_0 x + g\,z(x,y) + \alpha]; \quad \omega_0 = v_0\cos(\varphi_0),\ g = \tan(\varphi_0). \tag{1}$$

The digitizing object is $z = z(x,y)$. The spatial-carrier of the projected fringes is $v_0$. The fringe projector has a phase-sensitivity angle $\varphi_0$ with the $(x,y)$ plane. The angle $\alpha$ is the phase-shifting used in a phase-stepping algorithm (PSA) to demodulate $g\,z(x,y)$. Assuming telecentric illumination and imaging the spatial-carrier as imaged over the CCD plane is $\omega_0 = v_0\cos(\varphi_0)$, and its phase-sensitivity is $g = v_0\tan(\varphi_0)$. Finally note that the left fringe pattern $I_L(x,y,\alpha)$ has negative modulating phase $-g\,z(x,y)$, while the right fringe pattern $I_R(x,y,\alpha)$ have positive modulating phase $+g\,z(x,y)$.

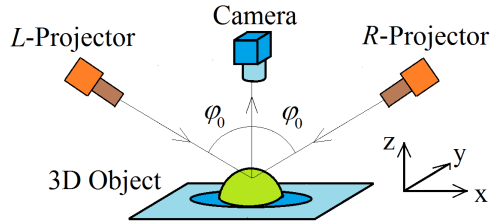

Figure 1. Co-phased fringe-projection digitalization of solids using two-projection perspectives and a single-camera reported by Servin et. al. [17].

The complex analytic signal from a 4-step phase-shifting algorithm (PSA with $\alpha = \pi/2$), from the left-projector is the following,

$$A_L(x,y)e^{ig\,z(x,y)} = I_L(0) + I_L(\pi/2)e^{i\pi/2} + I_L(\pi)e^{i\pi} + I_L(3\pi/2)e^{i3\pi/2}. \tag{2}$$

From the right-projection perspective the analytic signal is,

$$A_R(x,y)e^{-igz(x,y)} = I_R(0) + I_R(\pi/2)e^{i\pi/2} + I_R(\pi)e^{i\pi} + I_R(3\pi/2)e^{i3\pi/2}. \quad (3)$$

Note the minus sign on the phase-signal $gz(x,y)$ in Eq. (3). The two analytic signals in Eq. (2) and Eq. (3) have some complementary fringe contrast because they have different self-occluding shadows cast by the solid discontinuities; that is either $A_L(x,y) = 0$ or $A_R(x,y) = 0$. At places where self-occluding shadows occur due to the solid discontinuities the phase $gz(x,y)$ in Eq. (2) or Eq. (3) is not defined. But their co-phased sum is perfectly defined everywhere,

$$A_{L+R}(x,y)e^{igz(x,y)} = A_L(x,y)e^{ikg(x,y)} + \left[A_R(x,y)e^{-igz(x,y)}\right]^*,$$
$$A_{L+R}(x,y)e^{igz(x,y)} = \left[A_L(x,y) + A_R(x,y)\right]e^{igz(x,y)}. \quad (4)$$

Where $[.]^*$ denote the complex conjugate of the bracketed function. In this way using a co-phased 2 projectors, 1-camera (see Fig. 1) profilometer one overcomes the shadows cast by the object [17]. Remember, to avoid the solid's shadows we must have the magnitude sum in Eq. (4) $|A_{L+R}(x,y)| \gg \varepsilon$; being $\varepsilon$ a small real number.

## 3. Fringe-projection 360-degree profilometer using 1-projector and 1-camera

A previously reported 360-degree 1-projection, 1-camera profilometer for continuous solids is shown in Fig. 2(a) [18]. Let us assume that the digitizing solid may be represented by a single-valued continuous function $\rho = \rho(z,\varphi)$ in cylindrical coordinates,

$$\rho = \rho(\varphi, z); \quad \rho(\varphi, z) \in C^1; \quad \rho \in [0, R], \quad z \in [-z_0, z_0], \quad \varphi \in [0, 2\pi). \quad (5)$$

The digitizing solid shown are expected to fit inside a cylinder with radius $R$ and height $2L$.

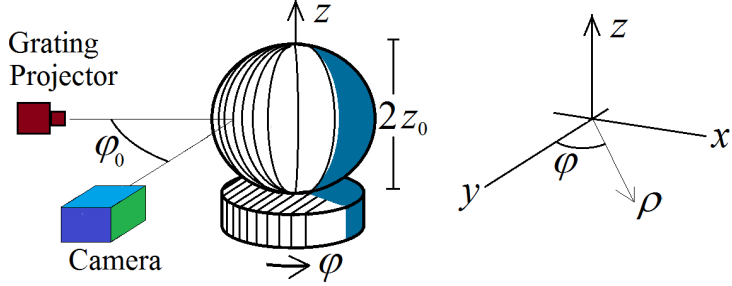

Fig. 2. Panel (a) shows our previous 360-degree profilometer set-up when a single linear grating projector and a single camera are used [18]. The linear grating and the camera are aimed towards the solid orthogonally to the $z$ direction and the phase sensitivity of this profilometer is proportional to $\tan(\varphi_0)$. Both the projector and the camera have their optical axis lying at the plane $(x,y,z=0)$.

The analyzing solid (the sphere in Fig. 2) is located over an angular-stepping turntable which can rotate a full revolution $\varphi \in [0, 2\pi)$ [18]. The number of turntable increments is $N$, so each rotation increment is $\Delta\varphi = 2\pi/N$. At each increment $n\Delta\varphi$, $n = \{0,1,\cdots,N-1\}$ the camera grabs the following fringe pattern over its sensing CCD array parallel to the $(x,z)$ plane,

$$I(x, z, n\Delta\varphi) = a(x, z, n\Delta\varphi) + b(x, z, n\Delta\varphi)\cos[\omega_0 x + g\,\rho(x, z, n\Delta\varphi)]. \quad (6)$$

Assuming telecentric illumination and imaging, the spatial frequency at the CCD plane is $\omega_0 = v_0 \cos(\varphi_0)$ and the profilometer sensitivity is $g = v_0 \tan(\varphi_0)$; being $v_0$ the spatial-carrier of the projected fringes. From the *N*-images in Eq. (6) we are only interested at the column pixels located at the center (*x*=0,z) or $I(0,z,n\Delta\varphi)$ for $n = \{0,1,\cdots,N-1\}$ [18],

$$I(0,z,n\Delta\varphi) = a(0,z,n\Delta\varphi) + b(0,z,n\Delta\varphi)\cos[\omega_0 0 + g\rho(0,z,n\Delta\varphi)]. \quad (7)$$

We may drop the "dummy" variable (*x*=0) to obtain,

$$I(z,n\Delta\varphi) = a(z,n\Delta\varphi) + b(z,n\Delta\varphi)\cos[g\rho(z,n\Delta\varphi)]. \quad (8)$$

Summing these *N* sampled centered pixel-columns, one digitize the entire object $\rho(z,\varphi)$ all around it (360-degrees) as,

$$I(z,\varphi) = \sum_{n=0}^{N-1} I(z,n\Delta\varphi)\delta(\varphi - n\Delta\varphi); \qquad z \in [-z_0, z_0], \varphi \in [0, 2\pi). \quad (9)$$

Being $\delta(\varphi - n\Delta\varphi)$ the angular-sampling Dirac delta function. Given that this fringe pattern (Eq. (9)) has no carrier ($\omega_0 x = \omega_0 \cdot 0 = 0$), the solid's phase $g\rho(z,\varphi)$ must be obtained using a temporal phase-shifting algorithm (PSA) [18].

## 4. Co-phased 2-projections 1-camera profilometer for 360-degree digitizing of solids

The co-phased 360-degree profilometry set-up is shown in Fig. 3. Our previous 1-projector 1-camera profilometer only permits to digitize continuous solids within the function space $\rho(z,\varphi) \in C^1$ [18]. In contrast using this co-phased 2-projections 1-camera profilometer it is now possible to digitize the wider function space of single-valued discontinuous 3D-surfaces $\rho = \rho(z,\varphi)$. So the continuity $\rho(z,\varphi) \in C^1$ limitation in [18] is removed.

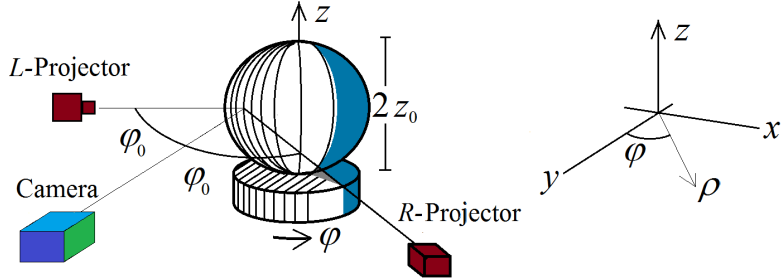

Fig. 3. Proposed co-phased 2-projectors 360-degree profilometer to digitize piecewise-continuous surfaces $\rho=\rho(z,\varphi)$. The projected linear gratings and the camera reside in the (*x,y,z=0*) plane. The phase-sensitivity of this co-phased profilometer is proportional to tan($\varphi_0$).

With 2-projections one eliminates the self-generated object shadows cast by the discontinuities (see Fig. 4) of the solid. The discontinuous 3D-surfaces that can be digitized with our co-phased profilometer belong to the single-valued discontinuous (piecewise-continuous) function space $\rho = \rho(\varphi, z)$; $\rho \in [0, R]$, $z \in [-z_0, z_0]$, $\varphi \in [0, 2\pi)$.

In Fig.4 we show a schematic cross-section of a discontinuous solid $\rho = \rho(z,\varphi)$ having a total of 6 discontinuities. Let us assume that we obtain 4-phase-shifted fringe patterns from Eq. (9). We label these fringe patterns as $I_L(z,\varphi,0)$, $I_L(z,\varphi,\pi/2)$, $I_L(z,\varphi,\pi)$ and $I_L(z,\varphi,3\pi/2)$ for the left-projector. Similarly we need 4-phase-shifted fringe patterns for the right-projector $I_R(z,\varphi,0)$, $I_R(z,\varphi,\pi/2)$, $I_R(z,\varphi,\pi)$ and $I_R(z,\varphi,3\pi/2)$. Using a 4-step PSA one can obtain the following analytic signal for the left fringe projector.

$$A_L(z,\varphi)e^{i\,g\rho(z,\varphi)} = I_L(0) + I_L(\pi/2)e^{i\pi/2} + I_L(\pi)e^{i\pi} + I_L(3\pi/2)e^{i3\pi/2}. \quad (10)$$

The coordinates $(z,\varphi)$ of the fringe patterns were omitted for clarity. Using the same 4-step PSA the analytic signal obtained for the right projector is,

$$A_R(z,\varphi)e^{-ig\rho(z,\varphi)} = I_R(0) + I_R(\pi/2)e^{i\pi/2} + I_R(\pi)e^{i\pi} + I_R(3\pi/2)e^{i3\pi/2}. \quad (11)$$

Note that the object phase $g\rho(z,\varphi)$ is negative in Eq. (11).

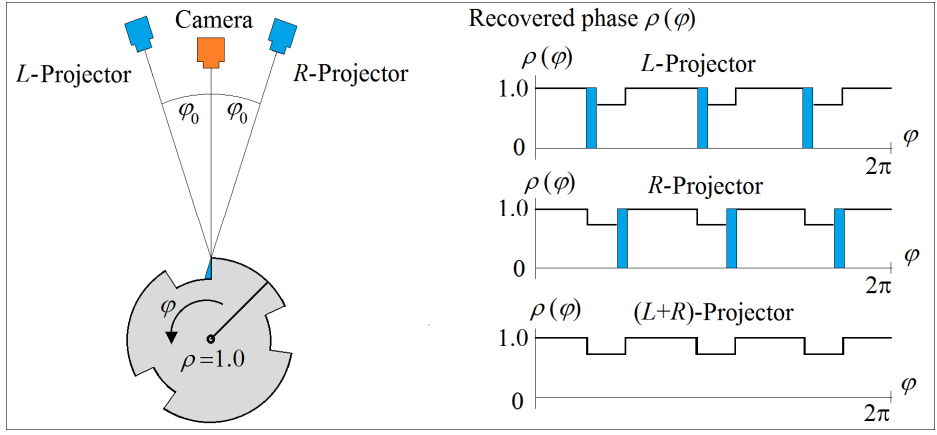

Fig. 4. Here we show the cross-section schematic of a dented solid $\rho=\rho(z,\varphi)$ having 6 discontinuities. In the specific rotation angle $\varphi$ shown, the *L*-projector cast no shadows over the discontinuity as seen from the CCD-camera. On the other hand the *R*-projector cast a complementary shadow at the discontinuity where the phase is not defined. The blue-zones in the graphs denote zero fringe-contrast where individual *L* or *R* phases are undefined. Co-phased summing the analytic signal of both projections the phase $g\rho(z,\varphi)$ is now well defined even at the discontinuity neighborhoods.

At the shadow discontinuities the analytic amplitudes drops to zero, *i.e.* $A_L(z,\varphi)=0$, or $A_R(z,\varphi)=0$. As a consequence the digitized phase $g\rho(z,\varphi)$ in Eq. (10) and Eq. (11) are not defined. But their co-phased sum will be defined everywhere,

$$A_{L+R}(z,\varphi)e^{ig\rho(z,\varphi)} = A_L(z,\varphi)e^{ig\rho(x,y)} + \left[A_R(z,\varphi)e^{-ig\rho(z,\varphi)}\right]^*,$$
$$A_{L+R}(z,\varphi)e^{ig\rho(z,\varphi)} = \left[A_L(z,\varphi) + A_R(z,\varphi)\right]e^{ig\rho(x,y)}; \quad z\in[-z_0,z_0], \varphi\in[0,2\pi). \quad (12)$$

The symbol $[\cdot]^*$ denotes the complex conjugate of the bracketed argument. These two analytic signals $A_L(z,\varphi)e^{ig\rho(z,\varphi)}$ and $A_R(z,\varphi)e^{ig\rho(z,\varphi)}$ are co-phased and have complementary shadows information because they have different self-occluding regions cast by the discontinuities in $\rho(z,\varphi)$. But their sum $A_{L+R}(z,\varphi)\exp[ig\rho(z,\varphi)]$ has no shadows.

A fundamental limitation of any 360-degree fringe-projection profilometer is when the solid's 3D-surface $\rho=\rho(z,\varphi)$ fails to be single-valued as shown in Fig. 5.

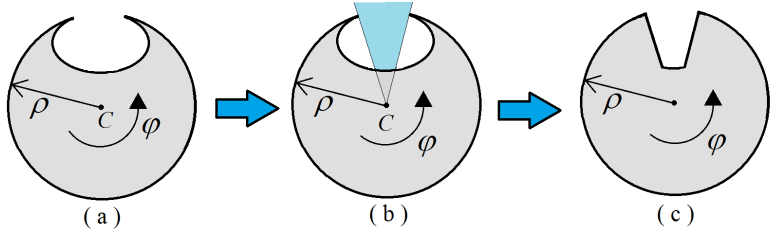

( a )        ( b )        ( c )

Fig. 5. Here we show a cross-section of a solid with a large shallow hole at plane $z=z_1$. The solid has its rotation center at point $C$. From the viewing perspective of the camera the hole in panel (a), is seen as the blue-wedge hole shown in panel (b). So this cross-section will be erroneously digitized as panel (c) shows.

Figure 5 shows a solid cross-section $\rho = \rho(z_1, \varphi)$ at plane $z = z_1$ containing a deep-shallow hole. As can be see this cross-section fails to be single-valued in some regions. The 3D digitalization will be "the visible cross-section" as seen from the camera marked as a blue-wedge in Fig. 5(b). The resulting erroneous digitized cross-section is shown in Fig. 5(c).

## 5. Experimental results

Here we demonstrate an experimental result of our 2-projector co-phased 360-degree profilometer by digitizing a highly discontinuous model of a human skull shown in Fig. 6.

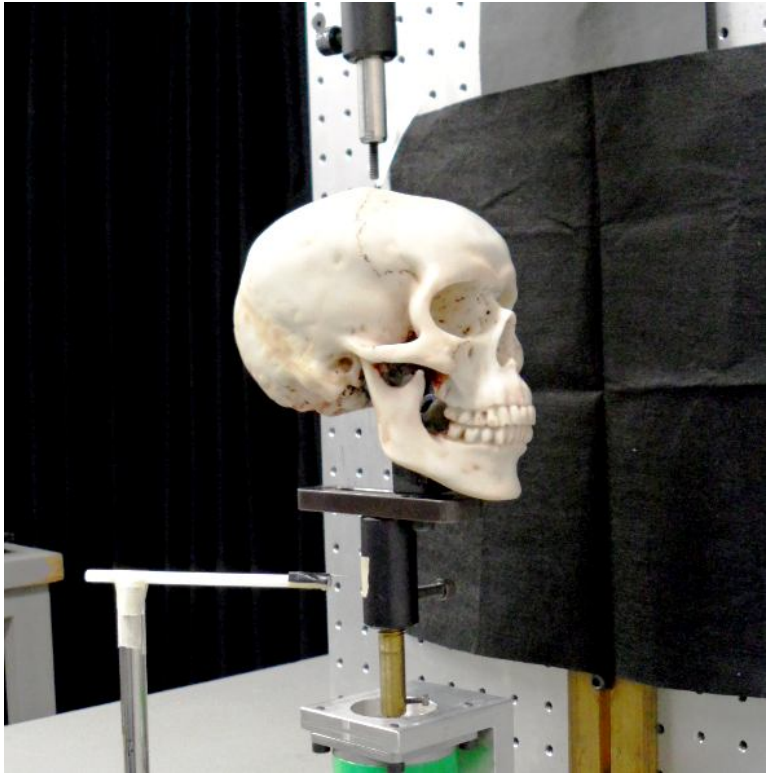

Fig. 6. Here we show the plastic model of a human skull that was chosen to demonstrate our co-phased 2-projector, 1-camera 360-degree profilometer. This model was chosen because its challenging discontinuities and deep holes generate many self-occluding shadows.

*5.1 The 360-degree fringe projection process.*

Figure 7 shows a plastic human skull to clearly see the advantages of using co-phased 2-projections 360-degree profilometry. The two images in Fig. 7(a) show the camera's view when only the left-projector illuminates the skull; this generates left-shadows at the eye basins. On the other hand, the last two images labeled as Fig. 7(b) are illuminated from the right-projector generating right-shadows. That is, the left-projection generates left-shadows while the right-projection generates right-shadows. By co-phasing summing the demodulated analytic signal from the right and left projections (Eq. (4)), the self-occluding skull's shadows are automatically eliminated. Also in Fig 7 we show the centered column line-pixel ($x=0,z$) of the projected fringes.

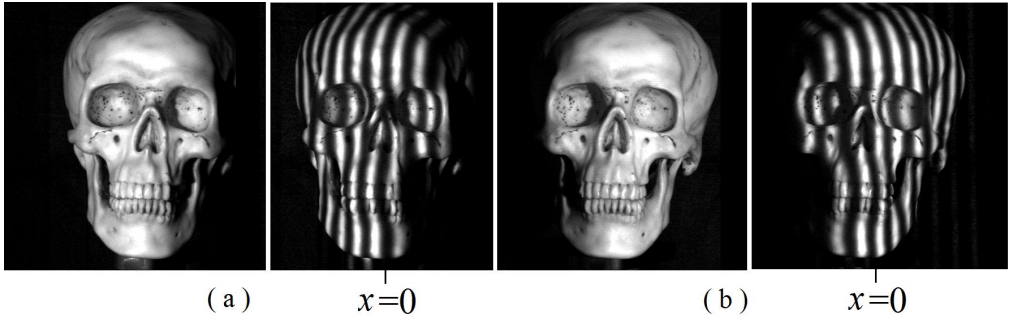

(a)  x=0  (b)  x=0

Fig. 7. Here we show the plastic human skull from the two projectors' perspectives. In the right hand side we show the left white-light and fringe projection, while the two last figures shows the shadows cast by the right projector.

Figure 8 shows the same left and right projections as viewed by the CCD camera when the skull is positioned at one of its sides. The first two images correspond to white-light and fringes from the right-projector. The last two images in Fig. 8(b) corresponds to white-light and fringes from the left-projector.

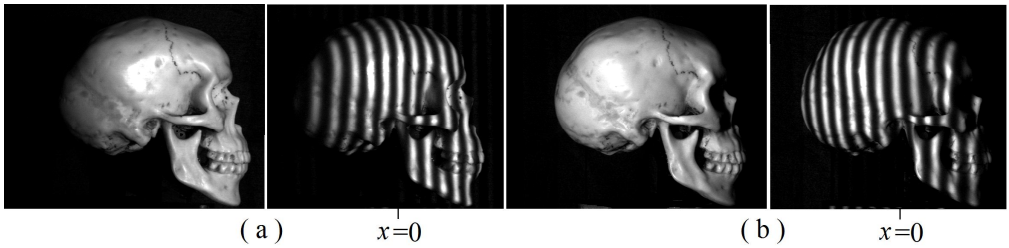

(a)  x=0  (b)  x=0

Fig. 8. Here we show the plastic skull from the orthogonal side with respect to Fig. 7. The two right hand side images show the left-projector illumination, while the two last images show the shadows cast by the right-projector.

Figure 9 shows the 4-phase shifted (closed-fringes) fringe-patterns used to phase demodulate the linear fringes projected over the $Skull(z,\rho)$. As Eq. (9) shows these 4 fringe-patterns have no carrier so their phase-demodulation must be done using a phase-shifting algorithm (PSA). We have used a 4-step least-squares phase-shifting algorithm with a phase-step of $\alpha = \pi/2$.

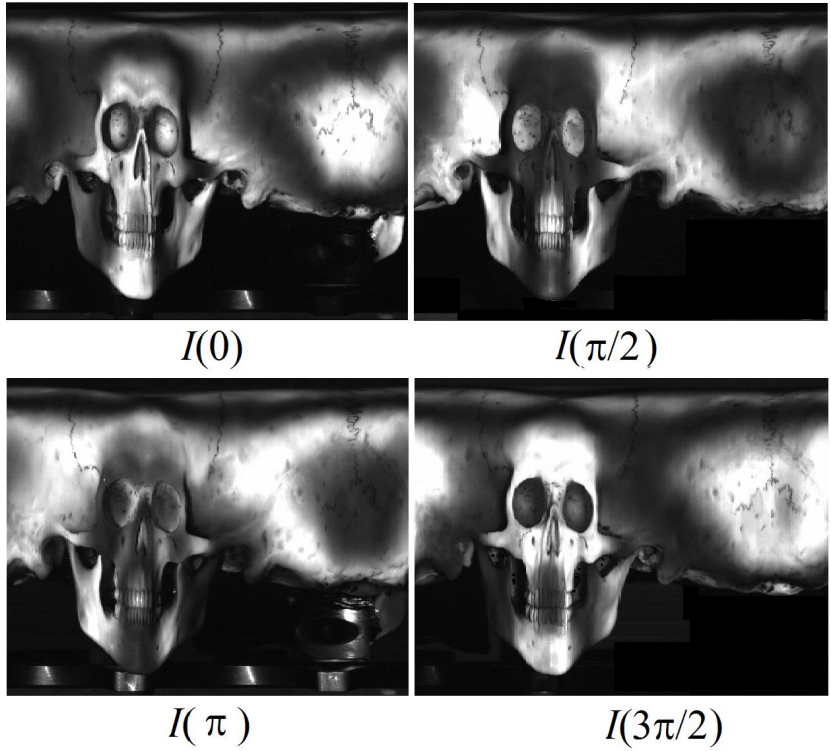

$I(0)$          $I(\pi/2)$

$I(\pi)$          $I(3\pi/2)$

Fig. 9. Here we show the set of 4-phase-shifting fringe patterns {$I(0)$, $I(\pi/2)$, $I(\pi)$, $I(3\pi/2)$} that were used to demodulate the projected fringes from the right-projector. A very similar set of 4 phase-shifted images was obtained from the illumination perspective of the left-projector.

*5.2 The 4-step phase-shifting phase demodulation and its unwrapping*

As seen previously, we have taken 4 phase-shifted fringe patterns for each projection perspective. So for the left-projector the 4 phase-shifted PSA used to obtain its demodulated analytic signal $A_L(z,\varphi)e^{ik\,Skull(z,\varphi)}$ is,

$$A_L(z,\varphi)e^{ig\,Skull(z,\varphi)} = I_L(0) + I_L(\pi/2)e^{i\pi/2} + I_L(\pi)e^{i\pi} + I_L(3\pi/2)e^{i3\pi/2}. \qquad (13)$$

The cylindrical spatial coordinates $(z,\varphi)$ was omitted for clarity. From the second fringe projection perspective the analytic signal is,

$$A_R(z,\varphi)e^{-ig\,Skull(z,\varphi)} = I_R(0) + I_R(\pi/2)e^{i\pi/2} + I_R(\pi)e^{i\pi} + I_R(3\pi/2)e^{i3\pi/2}. \qquad (14)$$

Note that the sign of the phase $g\,Skull(z,\varphi)$ has changed in Eq. (14). At the skull discontinuities the analytic amplitude drops to zero i.e. $A_L(z,\varphi) = 0$ or $A_R(z,\varphi) = 0$. As a consequence the phase $g\,Skull(z,\varphi)$ is not defined. But their co-phased addition is defined everywhere. Now let us sum these two co-phased analytic signals,

$$\begin{aligned} A_{L+R}(z,\varphi)e^{ig\,Skull(z,\varphi)} &= A_L(z,\varphi)e^{ig\,Skull(z,\varphi)} + \left[A_R(z,\varphi)e^{-ig\,Skull(z,\varphi)}\right]^*, \\ A_{L+R}(z,\varphi)e^{ig\,Skull(z,\varphi)} &= \left[A_L(z,\varphi) + A_R(z,\varphi)\right]e^{ig\,Skull(z,\varphi)}; \qquad z \in [-z_0, z_0], \varphi \in [0, 2\pi). \end{aligned} \qquad (15)$$

Where $[\cdot]^*$ denote the complex conjugate.

Now let us define 3 binary indicator function masks $m_L(z,\varphi)$, $m_R(z,\varphi)$ and $m_{L+R}(z,\varphi)$. These masks indicate the places at which the analytic signals have strong magnitude,

$$m_L(z,\varphi) = 1, \ if \ A_L(z,\varphi) > \varepsilon \ ; \quad otherwise \ m_L(z,\varphi) = 0,$$
$$m_R(z,\varphi) = 1, \ if \ A_R(z,\varphi) > \varepsilon \ ; \quad otherwise \ m_R(z,\varphi) = 0, \quad (16)$$
$$m_{L+R}(z,\varphi) = 1, \ if \ A_{L+R}(z,\varphi) > \varepsilon \ ; \quad otherwise \ m_{L+R}(z,\varphi) = 0.$$

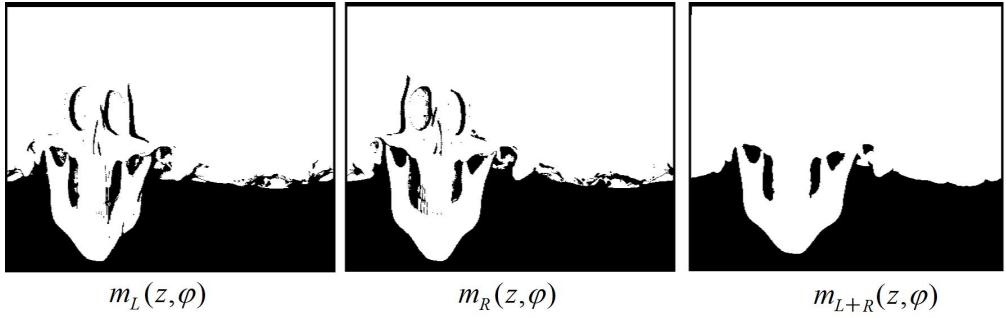

$m_L(z,\varphi)$          $m_R(z,\varphi)$          $m_{L+R}(z,\varphi)$

Fig. 10. The first image is the mask $m_L(z,\varphi)$ of valid fringe data (good fringe-contrast) from the left-projector. The second image is the mask $m_R(z,\varphi)$ of valid fringe data from the right-projector. Masks $m_L(z,\varphi)$ and $m_R(z,\varphi)$ show complementary shadow information. The third image shows the mask $m_{L+R}(z,\varphi)$ of strong analytic signal of the co-phased sum in Eq. (15).

We now use the mask sum $m_{L+R}(z,\varphi)$ to mask-out poor contrast amplitude regions as,

$$A'_{L+R}(z,\varphi) e^{ig\,Skull(z,\varphi)} = m_{L+R}(z,\varphi) [A_L(z,\varphi) + A_R(z,\varphi)] e^{ig\,Skull(z,\varphi)}. \quad (17)$$

The mask $m_{L+R}(z,\varphi)$ in Eq. (17) keeps only the regions where the co-phased analytic sum has strong magnitude to avoid undefined phase signal $g\,Skull(z,\varphi)$. Figure 11(a) shows the wrapped phase $Skull(z,\varphi)$ masked by $m_{L+R}(z,\varphi)$, while Fig. 11(b) shows the unwrapped phase $Skull(z,\varphi)$ also multiplied by $m_{L+R}(z,\varphi)$. The unwrapping process used was the standard line integration of wrapped phase differences; we did not use any sophisticated or complex phase unwrapper.

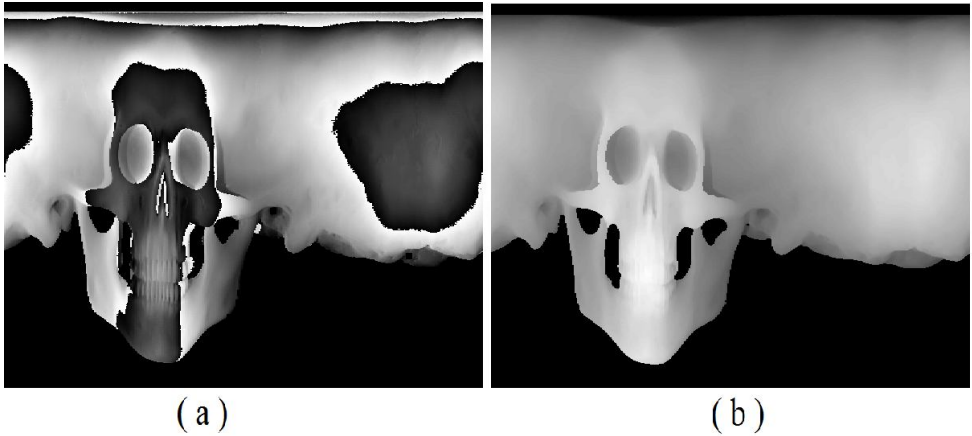

(a)          (b)

Fig. 11. Panel (a) shows the wrapped phase $Skull(z,\varphi)$ obtained by co-phased summing the two complex analytic signals from both projectors in Eq. (17). The wrapped phase $Skull(z,\varphi)$ is masked by $m_{L+R}(z,\varphi)$. Panel (b) shows the unwrapped $Skull(z,\varphi)$ obtained by standard line integration unwrapping of phase differences.

Figure 12 shows the unwrapped $Skull(z,\varphi)$ phase with the fringe amplitude $A'_{L+R}(z,\varphi)$ superimposed to better see the anatomical details of the skull model. These two images are shown to look how the phase $Skull(z,\varphi)$ looks in the flat digitized rendering space $(z,\varphi)$.

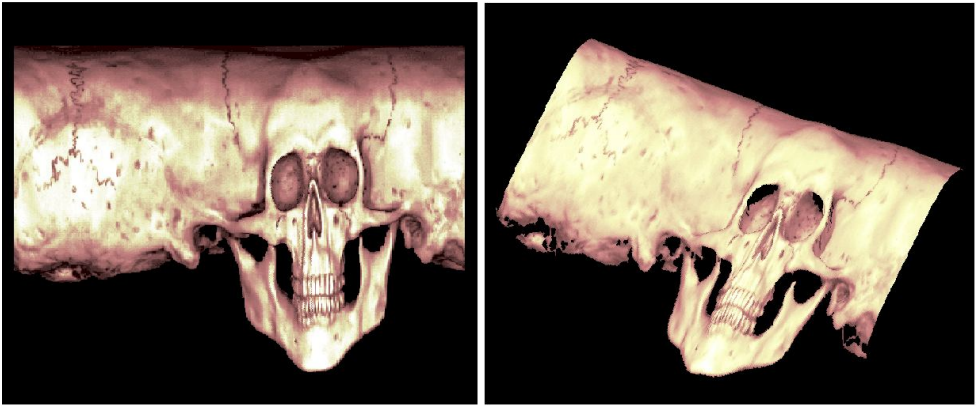

Fig. 12. Here we show the unwrapped skull phase $Skull(z,\varphi)$ yellow colored and with superimposed fringe magnitude $A'_{L+R}(z,\varphi)$ in Eq. (17). These two skull representations are masked by $m_{L+R}(z,\varphi)$..

Finally Fig. 13 shows the 3 rendering 3D-perspectives of the digitized plastic skull. The first image shows the $Skull(z,\varphi)$ phase in gray levels, while the other three panels shows 3 different perspectives of the digitized skull yellow-colored and with the fringe-magnitude texture to better visualize anatomical details.

As mentioned, the fringe patterns obtained have no spatial-carrier (see Fig. 9 and Eq. (9)) so we have to demodulate them using a 4-step phase-shifting algorithm (PSA). As a consequence of using a PSA one has the full spatial frequency spectrum to reconstruct fine details of the skull surface. In other words having close-fringes (base-band) fringe patterns (Fig. 9) one has the full spectral fringe digitized bandwidth to recover fine surface details. This high-frequency surface reconstruction (Fig. 11, Fig. 12 and Fig. 13) cannot be seen in previously published 360-degree profilometers which render very smooth surfaces $\rho = \rho(z,\varphi)$ [2-15].

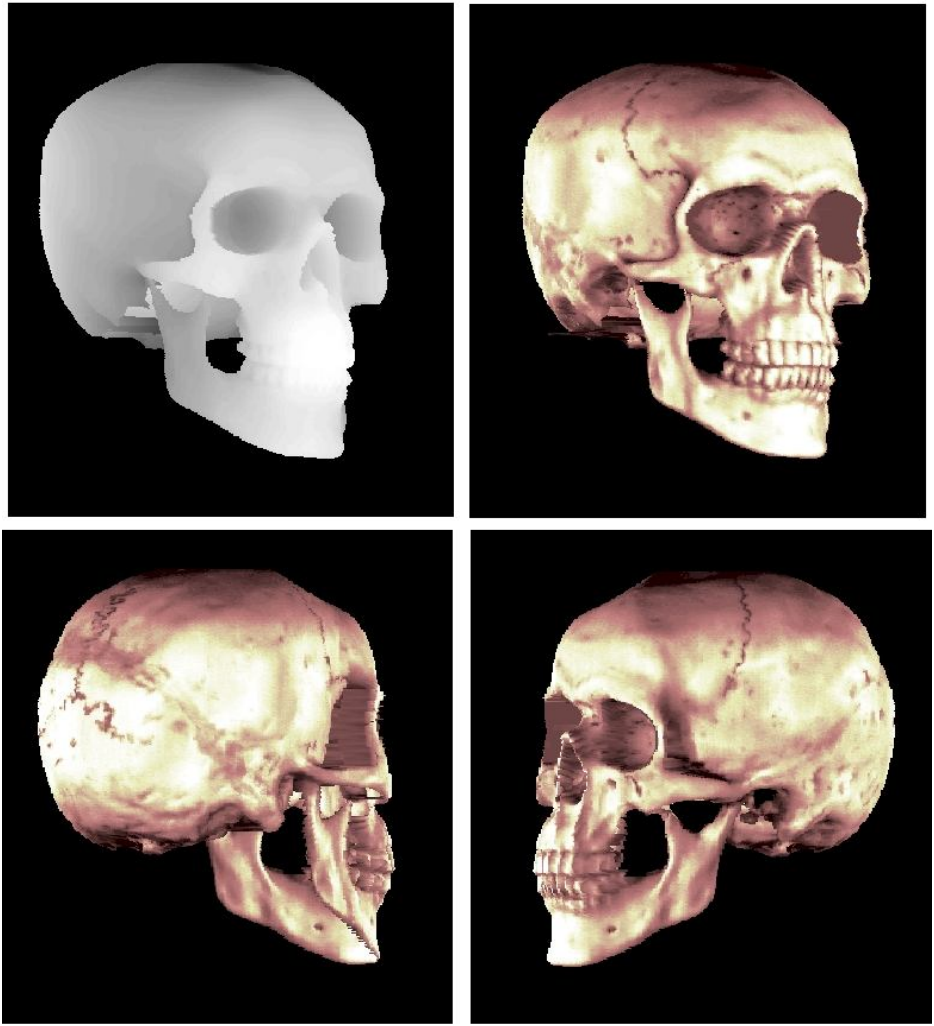

Fig. 13. The left image shows the plastic skull solid just in gray-level phase-values. The other three images show 3 perspectives of the digitized skull with superimposed fringe-contrast function $A'_{L+R}(z,\varphi)$ in Eq. (17) to clearly show the skull anatomical structures.

At the risk of becoming repetitive, we emphasize that, using base-band (closed-fringes) fringe patterns, allows one to reconstruct the digitized 3D-surfaces with the full theoretical spatial bandwidth that the raw digitized fringe data can hold.

### 6. Advantages and limitations of our co-phased profilometer for discontinuous solids

In Fig. 5 we have already highlighted one fundamental limitation of our co-phased profilometer where deep shallow holes cannot be rightly digitized at places where the surface $\rho = \rho(z,\varphi)$ becomes multi-valued.

As mentioned before, the digitizing surfaces $\rho = \rho(z,\varphi)$ that can be analyzed belong to the space of single-valued discontinuous (piecewise-continuous) functions. So in this paper we enlarge the functional space from continuous $\rho(z,\varphi) \in C^1$ in our previous work [18], to discontinuous functions. The space of discontinuous functions $\rho(z,\varphi)$ includes all convex solids as a proper subset. However the single-valued discontinuous function space fails to include more complex 3D-surfaces such as the boundary of a coffee-mug. That is because the space of discontinuous functions $\rho(z,\varphi)$ does not include multi-valued (in cylindrical coordinates) boundary surfaces such as a coffee-mug.

Another important limitation of this co-phased 360-degree profilometer is the maximum height of radial discontinuities that can be handled. To analyze this and for the reader's convenience let us re-write Eq. (9),

$$I(z,\varphi) = \sum_{n=0}^{N-1} \left\{ a(z, n\Delta\varphi) + b(z, n\Delta\varphi)\cos\left[g\rho(z, n\Delta\varphi)\right] \right\} \delta(\varphi - n\Delta\varphi). \qquad (18)$$

As this equation shows the sensitivity of this profilometer is given by $g = v_0 \tan(\varphi_0)$, so the maximum height discontinuity $\rho_{Max}(z,\varphi)$ that this profilometer can handle is,

$$\rho_{Max}(z,\varphi) < \left| \frac{2\pi}{v_0 \tan(\varphi_0)} \right|; \quad \forall\, z \in [-z_0, z_0],\ \forall\, \varphi \in [0, 2\pi). \qquad (19)$$

If higher discontinuities need to be handled one would need to use lower spatial frequency fringes, or use the temporal phase unwrapping for discontinuous objects [19].

**7. Conclusions**

Here we have presented a new experimental set-up for co-phased 360-degree profilometry which uses 2-projectors and 1-camera. This new co-phased set-up is capable of digitizing a wider set of 3D-surface functions $\rho = \rho(z,\varphi)$ than our previous 1-projector, 1-camera approach [18]. Our previous profilometer [18] was capable of digitizing solids only within the space of continuous functions $\rho(z,\varphi) \in C^1$. The herein presented co-phased profilometer is now capable of digitizing solids represented by single-valued discontinuous (piecewise-continuous) surfaces $\rho = \rho(z,\varphi)$. This profilometer can digitize 3D-surfaces without self-occluding shadows due to discontinuities.

We have chosen as test solid a plastic human skull for its sharp discontinuities, deep holes and high frequency surface details. As seen from the experimental results (Figs. 11, 12 and 13) we are capable of digitizing most of the discontinuous skull without self-generated shadows. That is because the projectors' perspectives complement each other eliminating the self-occluding discontinuities shadows. In contrast previous 1-projector, 1-camera 360-degree profilometers [2-18] generate self-occluding shadows at the solid's discontinuities. Finally using our profilometer one obtains base-band fringe patterns (Fig. 9 and Eq. (9)) keeping the full spatial 3D-surface bandwidth of the raw digitized fringe data. This allows us to visualize higher frequency surface details. In other words, closed-fringes pattern images, allows one to phase-shift demodulate solids with the highest possible digital bandwidth available from the raw fringe data.

**Acknowledgements**

The authors would like to acknowledge the financial support from project 177044 granted by the Mexican Consejo Nacional de Ciencia y Tecnologia (CONACYT).